\documentstyle[12pt]{article}

\newcommand{\beq}{\begin{equation}}
\newcommand{\eeq}{\end{equation}}
\newcommand{\bq}{\begin{quotation}}
\newcommand{\eq}{\end{quotation}}
\newcommand{\bc}{\begin{center}}
\newcommand{\ec}{\end{center}}

\def\s{\section}
\def\ss{\subsection}

\begin{document}

\title{Cosmology, Particles, and the Unity of Science}
\author{Henrik Zinkernagel\thanks{Consejo Superior Investigaciones
Cientificas, Instituto de Filosofia, Calle Pinar 25, Madrid,
Spain. e-mail: zink@ifs.csic.es}} 
\date{} 
\maketitle 

\vspace{-.5cm}
\begin{center}
\footnotesize 
To appear in {\em Studies in History and Philosophy of Modern Physics} 
\end{center}

\begin{quote}
\small
{\bf Abstract:} During the last three decades, there has been a growing
realization among physicists and cosmologists that the relation
between particle physics and cosmology may constitute yet another
successful example of the unity of science. However, there are
important conceptual problems in the unification of the two
disciplines, e.g. in connection with the cosmological constant and
the conjecture of inflation. The present article will outline some
of these problems, and argue that the victory for the unity of
science in the context of cosmology and particle physics is still
far from obvious. 
\end{quote}
\normalsize

\s{Cosmology and Philosophy} 

Given that cosmology has been part of human thought since the
beginning of intellectual history, it is somewhat surprising that
studies of the philosophy of scientific cosmology have been
remarkably few. Scientific cosmology dates back at least to the
beginning of this century, and although philosophers of science in
the 20th century have seen the advent of the theories of
relativity and quantum mechanics as the main events, the
all-important role of cosmology in the shaping of a scientific
`world-view' has not been done justice by this limited
philosophical attention. Perhaps this lack of interest has to do
with the fact that cosmology, for a large part of this century,
was not widely regarded as `real' science. Indeed, as
documented by Kragh \cite{kragh96}, up to at least the late
fifties there was no standard paradigm of cosmology, and more or
less philosophical arguments flourished in the debate over
steady-state theories vs. Big Bang alternatives. This situation
changed drastically with the 1965 discovery of the cosmic
background radiation, interpreted as the relic of the Big Bang,
which firmly established what was to be regarded as `normal
science' cosmology.\footnote{A detailed account of the history and
philosophy of cosmology before the microwave background radiation
discovery is given in North (1965) \cite{north65}.} 

The 1965 event, however, has by no means removed the need for
philosophical investigations of cosmology. In particular, the last
thirty years of interaction between cosmology and particle physics
have raised important questions regarding the relations between
philosophy, physics and cosmology. During this epoch steps have
been made towards the realization of a programme going back at least
to Eddington in the early 1930s of an intimate connection between
general relativity, the backbone theory of modern cosmology, and
quantum mechanics \cite{kilmister94}. Edward W. Kolb, a leading
theoretical astrophysicist, has expressed the relation between
cosmology and particle physics as follows:
\begin{quote}
In a very real sense the job of cosmology is to provide a canvas
upon which other fields of science, including particle physics,
can weave their individual threads into the tapestry of our
understanding of the Universe. Nowhere is the inherent unity of
science better illustrated than in the interplay between
cosmology, the study of the largest things in the Universe, and
particle physics, the study of the smallest things (\cite{kolb94} 
p.362). 
\end{quote}
It is clearly implied here that particle physics is {\em relevant}
for cosmology but the notion of interplay may also suggest a
stronger notion of unity between the two disciplines. Kolb points
out that `[t]he aim of modern cosmology is to understand the
origin and the large-scale structure of the Universe on the basis
of physical law' and there are those who take
the ultimate foundation of physical law to be grounded in particle
physics. Thus, it might be that particle physics could somehow be
used to {\em derive} cosmology and thereby function as the
`material' of which the cosmological `canvas' is 
made.\footnote{This would be in line with Einstein's declaration from
1918: `The supreme task of the physicist is to arrive at
those universal laws from which the cosmos can be built up by pure
deduction' \cite{einstein18}.}

Whatever the exact relation between cosmology and particle
physics, Kolb seems to be talking about unity of science in a
metaphysical sense of one common coherent picture of the world.
But while this conception of unity may be methodologically useful,
only a close inspection of the cosmology-particle physics
interplay can determine to what extent such a grand metaphysical
picture is in fact warranted. This state of affairs suggests a
central question for the philosophy of cosmology which will be
addressed in this article: What kind of unity obtains between
particle physics and cosmology -- for instance is particle physics
(merely) relevant to cosmology or can particle physics be used to
somehow derive cosmology? If an answer to this question can be
provided, it should be interesting also in connection with more
general unity claims in science -- not least, of course, if Kolb
is right that the interaction between cosmology and particle
physics provides the best illustration of the unity of science. 

In the following, I first turn to some general remarks on unity
and reductionism. Second, in order to illuminate the nature of
the cosmology-particle physics interplay, I review the
observational support for the Big Bang theory, and discuss some of
the conceptual issues which arise when particle physics is applied
to cosmology -- in particular the so-called cosmological constant
problem and the conjecture of inflation. Short discussions are
also included on other issues about the cosmology-particle physics
borderline such as the baryon asymmetry, dark matter, and the
problem of understanding the very notion of the early universe.
The final section summarizes and concludes.

\s{Unity With or Without Reductionism?}

The unity idea is often made explicit by referring to
reductionism. At the ontological level reductionism amounts to the
claim that entities and phenomena at a higher level can be reduced
to those of a lower level, for instance in a chain like classical
physics, atomic physics, particle physics, (string theory?).
Within physics, the discussion of reductionism is usually focused
on the notion of emergence -- the (ontological) question of
whether new and non-reducible phenomena exist at the higher levels
of description.\footnote{See e.g. Hacking (1996) \cite{hacking96}
and H\" uttemann (1998) \cite{huttemann98} for classification
schemes of various kinds of unity, and Cat (1998) \cite{cat98} for
a detailed discussion of the relationship between unity,
emergence, and (different notions of) reductionism in modern
physics.} For the discussion to follow, however, it will prove
more useful to briefly review a more semantic or epistemological
anti-reductionistic argument which can be associated with Bohr's
insistence on the necessity of classical physics in the
description of quantum phenomena. 

Recall that, for Bohr, there is no doubt that material objects
(e.g. a measuring apparatus) are composed of atoms. However, any
description of an atomic system which involves a determination of
its kinematical or dynamical variables must take the interaction
of the system with a suitable measuring apparatus into
consideration. Since a measuring apparatus, in order to yield
unambiguous experimental results, is necessarily described by
classical physics -- e.g. no superpositions of pointer readings --
the atomic (quantum) features of this apparatus must be ignored in
a description of a quantum experiment.\footnote{Bohr argued that
if the quantum features of the measuring apparatus, e.g. the
uncertainties in its momentum and position, are taken into
account, the `quantum' results of the measurements, e.g. an
interference pattern on a screen in the two slit experiment, are
destroyed, see e.g. \cite{bohr49}.} There is little doubt that
Bohr's views on the necessity of classical physics limits the
intelligibility of ontological reductionism: if quantum physics
does not make sense without classical physics, it is altogether
unclear what one should understand by reducing classical physics
to quantum physics.\footnote{Note that this is not in conflict
with the formalistic fact that quantum expressions may correspond
to classical expressions in certain limiting cases (e.g. when
$\hbar \rightarrow 0$). In fact, as quantum theories are often
constructed by `quantizing' classical theories, it is not
surprising that classical expressions may be formally derived from
quantum expressions in some appropriate limit. Furthermore, Bohr's
epistemological point would be unaffected even if one could
establish more concrete links between the quantum and classical
domain, e.g. through the notion of decoherence: if decoherence is
taken to imply an explanation of certain classical features from a
genuine quantum level, we would still be faced with the problem of
making sense of this level (see also below). And, as Faye has
noted, it is still far from clear what we should understand by a
`decoherence ontology' where `...a physical system in its
indeterministic plurality, with its probability distributions and
superposition, is the most fundamental entity in the world'
(translated from the Danish) \cite{faye96}. In any case,
decoherence -- as emphasized also by some of its spokesmen -- does
not by itself single out a particular interpretation of quantum
mechanics, see e.g. Zurek \cite{zurek91}.} 

Bohr's view on quantum mechanics implies that the classical
distinction between physical description and physical reality,
e.g. in the sense of the description mirroring reality, will no
longer do at the quantum level. The `no mirror' situation is just
another way of saying that the possibility of a full ontological
reduction of classical objects to quantum objects is excluded:
While atomic and sub-atomic objects exist independently of
measurements, the (space-time/energy-momentum) {\em properties} of
these entities do not, see e.g. \cite{bohr49}. If Bohr is right,
it thus becomes impossible to see classical phenomena as merely a
consequence of a deeper lying `quantum reality'. Rather classical
physics and quantum physics are both necessary for an account of
our knowledge of nature (see also Falkenburg (1998)
\cite{falkenburg98}). This argument against ontological {\em
reductionism} does not exclude ontological {\em unity} -- if the
latter is merely taken to mean that entities and phenomena of two
theories are interconnected (but not necessarily reducible to each
other). Indeed, Bohr emphasized both that, on the one hand,
classical physics is necessary to define quantum phenomena, and,
one the other hand, quantal laws are needed to explain the
stability of classical objects. 

As we shall see below, Bohr's arguments on the relationship
between the classical and quantum level are connected to the
discussion of the particle physics/cosmology interplay -- in
particular as concerns the question of whether cosmology may
eventually be derived from a deeper (quantum gravity) 
theory.

\s{The Cosmology-Particle Physics Interplay}

The interplay between particle physics ideas and cosmology
originated in work during the 1970s where it was realized that
phase transitions associated with particle physics might have
cosmological consequences (see e.g. the discussion in
\cite{rugh00}). Nevertheless, the exchange between cosmology and
microphysics in general is not just a product of modern particle
physics. Indeed Gamov, in some sense the creator of the prevailing
Big Bang paradigm, was led to his theory of the universe almost
exclusively through nuclear physics, in connection with an attempt
to account for the production of chemical elements -- which seemed
to require a very hot initial state (\cite{kragh96} p.81,391). In
the following subsections we will discuss to what extent this
relevance of microphysics to cosmology, and vice versa, is
observationally and/or experimentally supported when we move
from nuclear physics to particle physics. While this move may
appear as a straightforward change of scales (e.g. a move from
considering neutrons and protons to considering the constituents
of neutrons and protons) it involves the highly non-trivial change
of theoretical framework from quantum mechanics to quantum field
theory.\footnote{Although contemporary nuclear physics often
involves quantum field theory, the nuclear physics calculations by
Gamov and others, which were used to predict the light element
abundances, were not made in a quantum field theoretic framework;
see also below.} A distinctive feature of quantum field theory is
the peculiar concept of vacuum energy which, as we shall see
below, plays a key role in the cosmology-particle physics
interplay.

Intuitively, there seems to be a natural connection between
cosmology and particle physics: Experiments in particle physics
typically involve the study of reaction products from colliding
high energy particle beams. The higher the energies, the heavier
(and more exotic) the particles that can be produced in such
collisions. Since the Big Bang is believed to have involved
extremely high temperatures (energies), with the universe
gradually cooling down from a hot and dense initial state, the
early universe is portrayed as the ultimate particle physics
laboratory. In fact, since recent speculations of unification in
particle physics, e.g. that of grand unified theories and string
theories, refer to energy scales which are far outside the range
of particle accelerators on earth, the first tiny split second of
the history of the universe apparently becomes the only domain in
which such theories may be relevant. Although it is not easy to
obtain information of the very early universe -- for instance, the
Big Bang universe was optically opaque at times early than 300,000
years (see below), setting a limit to how far back in time
telescopes can probe -- it is hoped that some relics from this
early epoch can nevertheless be found in the cosmological data.
Thus, cosmological observations might be used to constrain, if not
test, speculative particle physics models. In turn, by addressing
the nature of matter and forces at extreme energy scales, such
particle physics models may supply explanations for theoretical
and observational puzzles in cosmology.\footnote{As the principal
setting for the conjectured unity between cosmology and particle
physics is the very early universe, we shall in this paper
disregard questions concerning the role of particle physics in
later astrophysical phenomena, for instance possible particle
physics explanations of the so-called solar neutrino problem.} 

If cosmology and particle physics are connected as outlined here,
there must be some theoretical bridges between descriptions of the
very small and the very large, or perhaps even a common framework
for both disciplines. In the remainder of this section we shall
focus primarily on two (closely related) theoretical bridges between
cosmology and particle physics -- the cosmological constant and
inflation -- which both turn out to throw doubt on the
cosmology-particle physics connection. As a further probe of the
relevance of particle physics to cosmology, we also briefly review
the status of two examples where particle physics models are
actively used in the search for answers to cosmological puzzles.
Finally, we briefly comment on the conceptual problems in an
attempt to provide a common framework for both cosmology and
particle physics. But since the relation between cosmology and
particle physics appears to rest on the assertion that the Big
Bang model is effectively correct, due to the extreme energy
conditions in the early universe, we first review the standard
textbook evidence for the Big Bang in order to see where particle
physics fits into the cosmological picture. 

\ss{Evidence for the Big Bang}

A central assumption in modern standard cosmology is that the
Friedmann equation, derived from general relativity and describing
the evolution of an expanding homogeneous and isotropic universe,
has been at least approximately valid almost throughout the
history of the universe. The faith in the Friedmann equation as
the correct description of the universe dates back to Hubble's
discovery in 1929 that other galaxies seem to move away from us,
indicating that the universe is indeed
expanding.\footnote{Actually, what Hubble found was a law relating
the distances to the galaxies to their observed redshift which was
subsequently interpreted, through the familiar Doppler effect, as
a universal expansion. We shall not here take up the interesting
issue of justifying this interpretation which has been contested
e.g. by adherents of the steady-state model.} This conclusion is
often reversed and extrapolated as far as logic permits with the
result that the universe must have been expanding from an initial
singularity. But, strictly speaking, as the light from the most
distant galaxies is supposed to have been emitted at a time when
the universe was a few billion years old, the observed Hubble
expansion does not say much about the very early
universe.\footnote{As we shall see later, however, even this
logical extrapolation breaks down at least when a time around
$10^{-43}$ seconds after the Big Bang is reached since classical
general relativity is supposed to break down at so-called Planck
scale energies where quantum gravity effects become important.
This Planck time is the result of a purely dimensional
consideration: it is the only time scale one can get by combining
Planck's constant, the velocity of light and Newton's
gravitational constant ($t_{planck}=\sqrt{G\hbar/c^5}= 10^{-43}
s$).} 

The time scale in the early universe is important, since the main
observational evidence for the Big Bang, apart from Hubble's
discovery, is the cosmic microwave background radiation, and the
abundances of the light elements in the universe, each of which
involves events at a specific point in time after the Big
Bang.\footnote{From the Friedmann equations for a radiation
dominated universe (in thermal equilibrium) it follows that
temperature in the early universe is related to time through $T
\sim 1/\sqrt{t}$. Note that this relation between time and temperature
is not valid in the very early universe for non-radiation
dominated epochs, e.g. during inflation.} The Cosmic Microwave
Background (CMB) radiation is believed to be a consequence of the
decoupling between matter and radiation which occurred around
$10^{13}$ seconds after the Big Bang (approximately 300,000 years).
The hypothesized primordial nucleo\-synthesis, on the basis of which
one can correctly predict the observed abundances of e.g. hydrogen
and helium, is estimated to have taken place when the universe was
roughly 1 second old (\cite{kolb93} p.109). Schematically the
evidence supporting the Big Bang, including the approximate time scale
which the evidence probes, can be represented as
follows:\footnote{The specification of exact years should be taken
with a grain of salt. As discussed in e.g. \cite{kragh96} Hubble
did not in 1929 interpret his results as is done nowadays,
nucleosynthesis was developed further after Gamov, Alpher and
Herman's work, and the CMB discovery by Penzias and Wilson had a
rich prehistory.} 

\begin{center}
\begin{tabular}{|l|l|l|} \hline
1929 & Hubble expansion  & $\sim$ 1 billion years after Big Bang\\
\hline                  
       & Gamov et al's prediction of & \\
1948 & light  element abundances & $\sim 1$ second after Big Bang \\
       &  from nucleosynthesis &  \\
\hline
1965 & CMB radiation & $\sim 300,000$ years after Big Bang\\
     \hline
\end{tabular}
\end{center}

But to reach an interaction point between particle physics and
cosmology one has to extrapolate the Friedmann equation much
further back than this evidence. Thus, the latest time where
physicists expect a pure particle physics (quantum field
theoretic) effect to have changed the cosmological situation is a
mere $10^{-5}$ seconds after the Big Bang; namely where a phase
transition associated with the theory of strong interactions is
conjectured to have caused the recombination of hadrons, e.g.
protons, from a primordial plasma of quarks and gluons. A phase
transition (spontaneous symmetry breaking) associated with the
electroweak theory is believed to have taken place still further
back in time at around $10^{-11}$ seconds after the Big
Bang.\footnote{It should be added that there is, so far, no direct
experimental evidence in the laboratory for either the quark-gluon
plasma or the electroweak phase transition (the latter is
associated with the so-called Higgs field which, in spite of
having being haunted for more than two decades, continues to
escape detection).} Earlier times than $10^{-11}$ seconds
(corresponding to higher energies) are well within the domain of
speculative particle physics; but it is clear that even the domain
of `standard' particle physics (from, say, 200 MeV to 300 GeV
corresponding to the range from $10^{-5}$ to $10^{-11}$ seconds
after the Big Bang) is {\em not} what is probed by light element
abundances -- and less by the CMB radiation, and still less by
light emitted from receding galaxies.\footnote{In fact, quantum
field theoretic corrections to e.g. the conversion rate between
neutrons and protons have been implemented in Big Bang
nucleosynthesis models which predict the abundance of primordial
Helium-4, see e.g. \cite{lopez98}. If one therefore had a
sufficiently accurate observational determination of this
primordial abundance, one might establish a more direct connection
between quantum field theoretical effects and cosmological
observations about the time of primordial nucleosynthesis.
However, it appears that recent observational determinations of
the primordial Helium-4 abundance are both controversial
(different methods yield inconsistent results) and too imprecise
to distinguish quantum field theoretic corrections in the
nucleosynthesis models; see e.g. \cite{sarkar99}.}

Although the observational pillars supporting the Big Bang thus
provide no direct support for the connection between cosmology and
particle physics, they may be used to constrain speculative
particle physics models. Examples of such `constraining'
cosmological data include: (i) the light element abundances, which
through their agreement with nucleosynthesis constrain the number
of neutrino species; and (ii) anisotropies in the CMB which can be
interpreted as relics from earlier epochs and therefore constrain
parameters e.g. in various inflation models (see e.g.
\cite{turner99}). But while cosmological data may provide
constraints on model building in particle physics, these
constraints can hardly constitute a {\em test} of the relationship
between the disciplines. For instance, the nucleosynthesis
constraint on the number of neutrino species arises since more
neutrino species would increase the primordial production rate of
Helium-4. Thus, some speculative
particle physics model which predicted more neutrino species than
allowed from the nucleosynthesis constraint would be in trouble
from the outset. However, as there is no theoretical prediction
(or explanation) of the number of neutrinos in {\em standard}
particle physics (the `Standard Model'), it seems reasonable to
maintain that the constraint from nucleosynthesis cannot be seen
as a cosmological test of a particle physics prediction. Of
course, since nobody would deny that nuclear physics is related to
particle physics, it would be foolish to assume that the Big Bang
nucleosynthesis model is correct but that no particle physics is
involved (in the finer details of the model). The argument made
here, however, is only that there is so far no direct evidence,
for instance from the confrontation between nucleosynthesis
predictions and observed abundances of the light elements, which
qualifies as a cosmological test of the particle physics before
(or at) the time of nucleosynthesis.

From this short discussion it seems in any case clear that the
kind of unity which particle physics and cosmology may possess
must be quite different from that of e.g. electricity and
magnetism -- which holds for our present epoch without any
recourse to a historical development, and which has direct and
manifold experimental and observational consequences. We now turn
from Big Bang evidence to some of the more conceptual problems to
be overcome in establishing a connection between cosmology and
particle physics. 

\ss{The Problem of the Cosmological Constant}

Arguably, the most important theoretical prerequisite for unity
between cosmology and particle physics will be a suitable
consistency between general relativity (GR) and quantum field
theory (QFT). But here, there arises an important problem known as
the problem of the cosmological constant.\footnote{The first part
of the present subsection is based on a manuscript, co-authored
with Svend E. Rugh, which provides a detailed analysis of the
historical and conceptual aspects of this problem, as well as the
physics on which it relies \cite{rugh00}. In the following, most
details of this analysis are omitted in an attempt to give an
overview of how the cosmological constant problem is related to
the question of unity between cosmology and particle physics.} In
essence the problem is that whereas the cosmological constant in
GR is experimentally known to be zero or close to zero, QFT
predicts an astronomically high value for this
quantity.\footnote{Recent observations on supernovae have
indicated that the cosmological constant in GR is small but
non-zero. In fact, one may distinguish three different
``cosmological constant problems": The observational question of
the value of the cosmological constant, the problem of the GR-QFT
clash dealt with here, and an `expected-scale' problem saying that
if the possible successor to the Standard Model is a quantum
theory of gravity characterized by the Planck mass $m_p$, one
would naively expect the observed cosmological constant to be of
order $m_p^4$.} The appearance of a `cosmological' constant in
particle physics is due to the fact that the vacuum energy in QFT
takes the form of the cosmological constant in Einstein's
equations. 

While the vacuum energy can be explained away in QFT since only
{\em differences} in energy matter, GR is sensitive to any {\em
absolute} value of this vacuum energy. Thus a huge vacuum energy
in GR would wrap up all of spacetime at macroscopic scales -- in
striking contrast to what is in fact observed. More technically,
Einstein's equations read

\begin{equation}
\label{Einstein1917} R_{\mu \nu} - \frac{1}{2} g_{\mu \nu} R -
\Lambda g_{\mu \nu} = - \frac{8 \pi G}{c^4} \; T_{\mu \nu}
\end{equation}
where $ R_{\mu \nu}$ and $R$ refer to the curvature of spacetime,
$g_{\mu \nu}$ is the metric, $T_{\mu \nu}$ the energy-momentum
tensor, $G$ the gravitational constant, and $c$ the speed of
light. 
A non-zero value of Einstein's cosmological constant would lead to
a curvature of `empty' space, i.e. when no matter or radiation is
present ($T_{\mu\nu} =0$), and it is therefore usually assumed that
$\Lambda$ receives contributions from any energy which might be
present in vacuum.\footnote{Einstein originally introduced his
`anti-gravity' constant to secure a static universe. Not least
after Hubble's observations, which indicated an expanding
universe, Einstein became very sceptical about the constant and is
reported to have called it `the biggest blunder of my life'; see
also \cite{ray90}.} In particular, the quantum expectation value
of the vacuum energy density ($<\rho_{Vac}>$) in particle physics,
which is composed of the vacuum energy of all the fields in QFT,
is supposed to act as such a contribution to the cosmological
constant: 
\begin{equation}
\label{Lambdaeff}
\Lambda = \Lambda _0 + \frac{8 \pi G}{c^4} < \rho_{vac} >
\end{equation}
where $\Lambda _0$ can be referred to as a `bare' cosmological
constant (representing any `non-standard' vacuum energy which would
curve empty space when no matter or radiation is present). Each of
the individual terms which build up the vacuum energy density in
QFT is estimated to be incredibly much larger, approximately
$40 - 120$ orders of magnitude, than the value of
the cosmological constant allowed from observation. The
discrepancy, in turn, can only be circumvented by assuming that
some mysterious cancellations take place in the total contribution
to the cosmological constant (either between the individual
contributions to QFT vacuum energy or between the terms in
eq.(\ref{Lambdaeff})) -- cancellations which, on our present
understanding of QFT and GR, seem highly implausible. 

Note that the introduction of {\em quantum} vacuum energy, which
rests on the notion of a fixed (Minkowski) spacetime background,
in a {\em classical} equation, which associate matter and energy
with a dynamical (non-Minkowski) spacetime, represents a
conceptual clash between two altogether different frameworks. It
is therefore not self-evident whether such a semi-classical
approach, which involves the difficult issue of treating quantum
fields in curved spacetime backgrounds, is reasonable for even
formulating a cosmological constant problem. In fact, as discussed
in detail in \cite{rugh00}, the notion of a cosmological constant
problem rests fundamentally on two assumptions -- both of which
can be questioned: (i) The QFT vacuum energy is physically real
(as in the standard QFT interpretation); and (ii) the
semi-classical approach, in which quantum energy acts as a
cosmological constant, is valid. As is evident from
eq.(\ref{Lambdaeff}), this last assumption implies a theoretical
bridge between GR and QFT. It can be argued that curved spacetime
effects on QFT are mostly small -- this is the case e.g. in our
local astrophysical neighbourhood where the gravitational field is
rather weak -- and hence that the vacuum energy is reasonably well
defined in the semi-classical context (this provides some
justification for (ii)). So the estimate of vacuum energy in
standard QFT appears in any case to be a severe obstacle to
compatibility between GR and QFT. Consequently, the cosmological
constant problem threatens the inter-theoretical consistency
between cosmology and particle physics. 

However, it can also be argued that the cosmological constant
problem, instead of being a problem for unity, points in the
direction of {\em further} unification in terms of some quantum
theory of gravity.\footnote{As the name suggests, quantum gravity
treats gravity in a quantum manner, but there exist a number of
different approaches for how to do that, see e.g. Butterfield and
Isham (1999) (\cite{butterfield99b} p.39 ff.). The `degree' of
unification attempted in the various approaches to quantum gravity
varies -- some approaches, like string theory, necessarily imply
unification of the gravitational force with the strong and
electroweak forces, whereas other approaches, e.g. the loop
quantum gravity programme, do not (\cite{butterfield99b}, p.66).}
In fact, the most promising solution to the cosmological constant
problem has sometimes been believed to lie in the framework of
quantum cosmology, in which quantum theory is applied to the
universe as a whole.\footnote{Although there is so far no
generally accepted solution to the cosmological constant problem,
a number of different ideas has been tried (see \cite{rugh00} for
a discussion and classification of the suggested solution types).
We shall here restrict attention to the quantum cosmology proposal
as this theory arguably provides the most radical attempt of
joining cosmology and quantum mechanics but it should be noted
that the most famous example of a quantum gravity, string theory,
is also plagued by the cosmological constant problem -- so far
with no cure in sight.} Quantum cosmology is usually based on the
so-called Wheeler-DeWitt equation for the `wave-function of the
universe', which results from the canonical approach to quantum
gravity where general relativity (the metric tensor) is quantized,
see e.g. (\cite{butterfield99b} p.65).\footnote{In contrast to
quantum cosmology, quantum gravity -- which may be characterized
as any attempt to join general relativity with quantum mechanics
(or quantum field theory) -- might also have applications other
than for early universe cosmology, e.g. black hole physics.} The
idea of a quantum cosmological solution to the cosmological
constant problem, which was first suggested by Hawking in 1984 and
further elaborated by Coleman in 1988, is to show -- by
introducing a new `quantum cosmological' gauge field or using
bizarre features of `quantum' spacetime known as wormholes --
that the value of the cosmological constant with very high
probability must be zero in our universe (see e.g.
\cite{weinberg89} p.20 ff. for discussion and references). We
shall not enter into any details of this proposal and its specific
problems but rather, in connection with the question of unity,
point out some of the general difficulties in using a quantum
cosmological framework for whatever purpose. 

Although no consistent quantum theory of gravity, and therefore no
consistent quantum cosmology, exist as yet, one can give some more
or less heuristic arguments in favour of why such theories are
eventually to be expected. For instance, it may appear strange
that gravity, as a force acting on matter which can otherwise be
described quantum mechanically, should not itself be quantum in
nature. A stronger argument along similar lines has been given by
a leading quantum cosmologist, Hartle, who
phrases his trust in a quantum theory of gravity in an almost
logical, if slightly paradoxical, form: 

\begin{quote}
It is an inescapable inference from the physics of the last sixty years
that we live in a quantum mechanical universe -- a world in which the
basic laws of physics conform to that framework for prediction we
call quantum mechanics. If this inference is correct, then there must
be a description of the universe as a whole and everything in it in
quantum mechanical terms. (\cite{hartle91} p.67)
\end{quote}
In the second sentence, Hartle makes two claims: 1) that there
must be a quantum description of the universe as a whole, and 2)
that it must be possible to describe everything in the universe in
quantum mechanical terms. The first of these claims is a claim of
the necessity of quantum cosmology; the second claims the
necessity of quantum gravity -- insofar as Hartle takes it for
granted that everything, besides gravity, has already been
described by quantum mechanics. Although, as indicated above,
these enterprises are closely related, we shall here just comment
on quantum cosmology, and return later to some remarks on the
possibility and necessity of quantum gravity. 

First, in spite of Hartle's historical `evidence', it is
questionable whether there is an `inescapable inference' from the
spectacular successes that quantum mechanics has had in the
microscopic domain to a statement that the whole universe is
genuinely quantum mechanical -- even taking into account that
quantum theory can be applied to some macroscopic systems such as
superconductors.\footnote{That this inference may not be
straightforward seems to be implicit also in Halliwell's
non-technical introduction to quantum cosmology
\cite{halliwell93}, see especially p.488. For more reservations on
the universal validity of quantum mechanics, see e.g. Hendry
(1998) \cite{hendry98}.} For one thing, insofar as the concept of
a `wave-function of the universe' is at all a sensible concept,
its governing Wheeler-DeWitt equation is ``horribly ill-defined in
any exact mathematical sense" (see e.g. \cite{butterfield99b}
p.65).

Another peculiarity of quantum cosmology, related to the (lack
of?) meaning of the wave function, is the absence of a precise
notion of measurement due to the lack of a clear distinction
between the observer and the system to be observed: contrary to
standard quantum mechanical systems there is obviously no
`external' system to the universe! In the discussion that follows
Hartle's quote above, he indicates that quantum cosmology
therefore poses a problem for the `Copenhagen' interpretation of
quantum mechanics (\cite{hartle91} p.67).\footnote{Quantum
cosmologists often invoke Everettian views on quantum mechanics
backed by the notion of decoherence, see e.g. \cite{kiefer99} p.10
and \cite{butterfield99a} p.38.} Nevertheless, one could also
argue the other way around: If Bohr's insistence of the necessity
of a classical framework for understanding quantum mechanics is
correct, then quantum cosmologists, not Bohr, have a problem. A
resolution of this issue depends on which, if any, of the
following is true: 1) the interpretation of quantum mechanics is a
matter of choice to fit the needs of the situation; or 2) one can
give general arguments against certain interpretations; or 3)
there are arguments of principle to single out one
interpretation.\footnote{See Faye \cite{faye96} for a discussion
of problems with so-called realistic interpretations of quantum
mechanics, and P. Zinkernagel \cite{zinkernagel62} for an account
of how the necessity of a classical framework can be understood in
a broader philosophical framework of `conditions for
description'.} Although this is not the place to go into this much
debated subject, it seems clear that Bohr's arguments, briefly
discussed in the previous section, demonstrate that the notion of
a `quantum reality' implied by the wave-function of the universe
might very well be unintelligible (see also below). In any case,
it is clear that discussions of quantum cosmology, for instance in
connection with the cosmological constant problem, take us right
back into the philosophical battlefield of interpreting quantum
mechanics.

Finally, even if we by-pass these conceptual issues for a moment,
the chances of actually observing quantum cosmological effects,
characterized by the Planck scale, seem remote. This situation is
illustrated by the fact that an extrapolation from the supposed
time elapsed from the Big Bang until nucleosynthesis (1 sec) back
to the Planck time ($10^{-43}$ sec) is considerably longer on a
logarithmic scale than an extrapolation from the present ($\sim
10^{17}$ sec) back to the time of nucleosynthesis! In this
connection, it should be noted that chances for testing aspects of
quantum cosmology may be further weakened by, ironically, another
aspect of early universe speculation, namely inflation. This is
because an inflationary period may well wipe out all traces of a
pre-inflationary (quantum cosmological?) epoch, see e.g.
(\cite{kolb93} p.458).

\ss{Inflation}

Inflation is short-hand for the idea that the expansion of the
universe at an early stage in its history proceeded extremely
fast. In turn, such a period of rapid expansion may provide a
solution to some conceptual problems in standard Big Bang
cosmology (see below). Originally, the idea of inflation was
conceived of as a consequence of the vacuum energy associated with
a spontaneous symmetry breaking in the Grand Unified Theory (GUT),
according to which the strong force and the electroweak force will
be unified at energies above $10^{14}$ Gev (in the simplest
version of the theory). This energy corresponds to a time $\sim
10^{-34}$ sec after the Big Bang -- and thus inflation was a clear
candidate for a solid contact point between cosmology and particle
physics.\footnote{The conversion between time and energy is found
from the Friedmann equation using the curious equation of state
for a vacuum energy dominated universe, $(energy \,\, density)= -
(pressure)$, see e.g. (\cite{kolb93} p.270 ff). The negative
pressure associated with positive vacuum energy density also
explains how that latter gives rise to a repulsive, anti-gravity,
effect, see e.g. \cite{rugh00}.} One way in which inflation has
been imagined is that during a short `inflation' period, lasting
some $10^{-32}$ seconds, the huge vacuum energy associated with a
GUT symmetry breaking field effectively acts as a cosmological
constant. This anti-gravitational vacuum energy boosts the
expansion of the universe exponentially until the vacuum energy is
finally converted into heat and the universe enters the standard
(slowly expanding) epoch.

Since vacuum energy is the source of almost all inflationary
scenarios, there is an explicit connection between the
cosmological constant (problem) and inflation: the vacuum energy
of some specific field was ``good" at early times -- inflation
runs on it so to speak -- but vacuum energy in general is not so
good now, for we know it must be very small or zero. This
connection is further illuminated since inflation, just like we
noted above for the cosmological constant problem, is cast in a
semi-classical framework: Due to its (anti-)gravitational effect,
quantum vacuum energy is driving a classical general relativistic
expansion of the universe. In this way inflation makes the
cosmological constant problem even more disturbing, and the latter
has therefore raised some critical voices on the inflation
speculations:

\begin{quote}
It seems difficult to understand the cosmological constant as a vacuum
energy density now, and it is therefore naive to exploit it this way
during the first $10^{-35}$ seconds of the universe 
(Boerner 1988) \cite{boerner88}.
\end{quote}
Also cosmologists with a more optimistic attitude to the idea of
inflation acknow\-ledge that the connection between the cosmological
constant problem and inflation may turn out to be fatal. Thus Kolb
and Turner writes:

\begin{quote}
...that puzzle [the cosmological constant problem] is a potential
Achilles heel for the scenario: If some grand principle should be
discovered that renders vacuum energy {\em at all times} impotent,
inflation would lose the ultimate source of its power (\cite{kolb93} 
p. 314).\footnote{Attempts have been made to solve the
cosmological constant problem in the context of quantum gravity 
and at the same time provide a mechanism for inflation, see 
\cite{brandenberger99} for discussion and references. Given the 
provisional nature of quantum gravity, however, these attempts
are, at best, incomplete.}
\end{quote}

Even if vacuum energy based inflation might seem to be in trouble
as a reasonable theory from the outset, the idea is claimed to
have provided cosmology with possible solutions to the horizon,
the flatness, and the monopole problems.\footnote{These problems
concern, respectively, the question of why the background
radiation is isotropic, why the observed universe is Euclidean to
a very high degree, and why no monopoles are observed (these
monopoles are objects which are expected on GUT grounds). See
Earman and Moster\'\i n (1999) for a critical analysis of
inflation and its merits as regards solving these problems
\cite{earman99}.} However, as a theory covering only the first
split second of the history of the universe, inflation's epistemic
status is problematic (even if not as problematic as that of
quantum cosmology). For instance, Narlikar noted in 1987 that
`The absence of monopoles, domain walls, seeds of galaxies, and
the rest is confidently taken to support a number of hypotheses
regarding the unification theory and the very early universe.
Nowhere else in physics have null observations been used to claim
so much that cannot be otherwise tested' \cite{narlikar87}. Now,
Narlikar is one of the few who still defends a competing world
view -- that of a `quasi' steady state model, a world without a
beginning in a Big Bang -- and so one might be hesitant about
accepting his challenge to inflation \cite{burbidge99}. Regardless
of world view preferences, however, Narlikar is surely right that
it is not particularly impressive if the observational `evidence'
of a theory is taken to consist of what is {\em not}
seen.

But there might of course be more than `null observations' which
could support inflation. To this end, it is often emphasized that
a generic prediction of inflation is an almost scale invariant
spectrum of the anisotropies in the microwave background
radiation.\footnote{For more discussion of generic predictions of
inflation, e.g. that of spatial flatness, see \cite{earman99}.}
Since this is actually what is seen in the data, the anisotropy
spectrum is a potential sign of inflation, although other cosmic
scenarios may lead to similar effects (\cite{brandenberger99}
p.19). This emphasis on anisotropies in the microwave background
radiation, however, can be a double-edged sword: For instance, all
GUT based inflation models generically predict the amplitudes of
these anisotropies to be far higher than microwave background data
allow -- and, in general, `...arranging for acceptable density
perturbations [directly linked to the anisotropies] is the most
restrictive constraint on inflationary potentials (and thereby the
underlying particle physics models)' (\cite{kolb93} p.288 and
295). The observational status of inflation is furthermore
complicated by the fact that the inflation industry has produced a
vast number of models which easily accommodate a wide variety of
observational situations:

\begin{quote}
The trouble is that inflation is a paradigm rather than a model
[...] a reasonable first guess is that at present there are  
around one hundred different models on the market, all consistent
(at least more or less) with present observational data.
(Liddle 1999 \cite{liddle99})
\end{quote} 

\noindent
Although it is therefore difficult to either support or rule out
the inflation idea by means of observations, planned detailed
studies of the microwave background in the near future are
expected to cast more light on the situation.\footnote{Recent
results from the balloon experiments Boomerang and Maxima seem to
be consistent with a spatially flat universe (which is predicted
by most inflation models) \cite{jaffe00}.} 

Worse than the observational issue, perhaps, for the relation
between cosmology and particle physics, are the developments in
inflation models which have shown that such models are in no
obvious way related to the spontaneous symmetry breaking of a
particle physics model. Already in 1983, Linde suggested a
so-called `chaotic' version of inflation which operates with a
very simple potential to drive the inflationary period (see e.g.
\cite{linde90b} p.66). While the simplicity of the inflation field
(the `inflaton') in Linde's model does not exclude some particle
physics origin of it being found, Linde made no attempt to link
his inflaton to particle physics. The chaotic inflation scenario
thus demonstrated, in contrast to the spirit of Guth's original
1981 proposal, that inflation by no means has to have a particle
physics origin. In accordance with his view on unity, however,
Kolb complains that although models such as chaotic inflation
realize the `true essence' of inflation without complicating
details of underlying physics, chaotic inflation `...is not
``realistic" in the sense that no one would accept the existence
of a scalar field whose sole purpose is to make inflation
simple...' (\cite{kolb94} p.32). And, indeed, while Linde's
proposal is still an area of active research, there is also an
ongoing tradition among imaginative physicists in search of a
basis for inflation in speculative particle physics models (see
e.g. \cite{brandenberger99}).

\ss{Other Problems on the Borderline?} 

In this subsection I briefly mention a few other topics on the
borderline between cosmology and particle physics. Although the
issues mentioned below do stimulate activity in both fields, they
do not unambiguously speak in favour of their being intimately
related. The following list is not intended to be exhaustive but
it illustrates that there are more puzzles on the way to unity
than `just' the cosmological constant and inflation. 

\ \\
{\em Baryon-asymmetry}. 
Observations of the universe indicate that the amount of
anti-matter (anti-baryons) is negligible in comparison to matter
(baryons), so that the universe seems to be baryon-antibaryon
asymmetric (see e.g. \cite{kolb93} p.158). Moreover, it turns out
that the correct predictions of the light element abundances in
the universe from the theory of nucleosynthesis depend directly on
the parameter $\eta$ which characterizes the baryon
asymmetry.\footnote{$\eta$ is defined as the ratio between the
excess of baryons over anti-baryons to photons and is supposed to
have remained roughly constant since the time of primordial
nucleosynthesis. Direct observations of baryonic matter in the
universe only account for one third of the value of $\eta$
required by successful primordial nucleosynthesis, but the
discrepancy is thought to be due to the presence of large
quantities of dark or invisible baryonic matter, see
(\cite{dolgov97} p.3) and also below.} 

On the other hand, if a particle physics model is to generate
such a surplus of baryons over anti-baryons; the obvious, though not
the only, condition is that the usual rule of baryon charge
conservation must be violated.\footnote{Sakharov (1967)
\cite{sakharov67} was the first to discuss the necessary features
of a particle physics model to generate the baryon asymmetry.} And
although processes where baryon charge is not conserved have
never been seen experimentally, physicists began to consider them
seriously with the advent of grand unified theories in the 1970s
in which such processes appear naturally \cite{dolgov97}. However,
simple versions of GUTs are now ruled out precisely because baryon
charge non-conservation processes, in particular the decay of the
proton, have still not been seen in spite of intensive searches
(the simple GUTs predict a lifetime of the proton which is incompatible
with lower limits from experiments). Nevertheless, for some
cosmologists, the observational fact of a baryon asymmetry is still
taken to support the cosmology/particle physics relationship: 

\begin{quote}
The baryogenesis scenario is one of the great triumphs of 
particle cosmology, and in the absence of direct evidence
for proton decay, baryogenesis may provide the strongest,
albeit indirect, evidence for some kind of unification of
the quarks and leptons [as in GUTs]. (Kolb 1993 
\cite{kolb93} p.158)
\end{quote}
But apart from the fact that baryogenesis has been proposed
in non-GUT scenarios, it is not clear that the asymmetry is in
particular need of an explanation from a general relativity based
cosmological point of view in the first place. 
Only if the quantum mechanical fact
that each particle has an anti-particle is taken into account, and
{\em if} it is assumed that quantum mechanics rules in the very
early universe (whatever that means), does the necessity of an
explanation arise. Indeed, to seek an explanation of the baryon
asymmetry is connected with the assumption that particle physics
is somehow the place to find cosmological answers, as noted in a
recent review of electroweak baryogenesis:\footnote{The idea of
electroweak baryogenesis, which is currently the most popular
scenario for generating the baryon asymmetry, is to explain the baryon
charge non-conservation on electroweak energy scales, through
so-called non-perturbative processes in the electroweak theory.
However, this scenario for baryogenesis is only viable when
non-tested extensions of the electroweak model, for instance
supersymmetry (a symmetry between fermions and bosons), is included 
in the description \cite{trodden98}.}

\begin{quote} We {\em could} just accept this [the baryon
asymmetry] as an input
parameter for the evolution of the universe. However it is part of
the philosophy of modern cosmology to seek an explanation for the
required value of $\eta$ using quantum field theories of
elementary particles in the early universe. (Trodden 1998
\cite{trodden98})
\end{quote} 
This is obviously a statement of choice but it is clear that once
the cosmological observed baryon asymmetry is supposed to be
explained by particle physics, the hitherto lack of generally
accepted scenarios based on experimentally tested ideas makes it
hard to judge whether the baryon asymmetry is good or bad news for
the connection between cosmology and particle physics.

\ \\ 
{\em Dark matter}. 
Like the baryon asymmetry, this issue has its origin in
astronomical observations. The hypothesis of dark matter arises as
a response to the problem of accounting for various astronomical
phenomena which seem to require much more mass than is observed. A
well-known example is the case of observed rotation curves for
galaxies: the estimated mass of luminous matter in the galaxies
cannot account for the observed rotation velocities of objects far
from the galaxy center (see e.g. \cite{kolb93} p.16-19). As
mentioned above, some of this dark matter must be baryonic, e.g.
in the form of white dwarfs, neutron stars, diffuse warm gas or
even black holes, to be consistent with the predictions about
nucleosynthesis. From comparisons between the primordial
nucleosynthesis scenario and observed abundances of e.g.
deuterium, it follows that the total amount of baryonic matter in
the universe is $\Omega _{baryon} \sim 0.03$, but since estimates
on the {\em total} amount of matter in the universe (e.g. from
studies of the density and properties of galaxy clusters
\cite{turner99}) suggest that $\Omega _{M} \sim 0.2-0.4$, one also
needs non-baryonic dark matter components.\footnote{We have here
used the convenient way of expressing the amount of a certain type
$x$ of mass or energy in the universe through the dimensionless
parameter $\Omega _{x}$. This is defined by $\Omega _x = \rho _x /
\rho _c$, where $\rho _x$ is the mass/energy density of $x$, and
$\rho _c$ corresponds to the critical mass/energy density in the
universe which distinguishes open, flat, and closed universes
(i.e. if $\rho _{total} = \rho _c$ and thus $\Omega _{total} =1$,
then the universe is flat).} Nevertheless, precise estimates of
the matter content in the universe are difficult to make, and
there are still those who think that all matter (dark and
luminous) may, in fact, be baryonic, see e.g. \cite{bosma98} and
(\cite{kolb93} p.369ff). 

Particle physics has a number of candidate models for what this
dark matter could be. Popular models include those of axions,
supersymmetric particles or massive neutrinos -- all of which are
characterized by the fact that they go beyond standard particle
physics and they refer to particles which are believed to have
been produced in the early universe. From observational
constraints it seems that massive neutrinos, denoted as `hot' dark
matter, can at most account for a small fraction of dark matter in
the universe. This suggests that `cold' dark matter, for instance
axions or supersymmetric particles, is the dominant component of
this invisible matter (see e.g. Turner and Tyson
\cite{turner99}).

While Turner and Tyson take observations to be a suggestive
pointer to the existence of nonbaryonic dark matter, another
leading cosmologist has this to say on the issue:
\begin{quote}
In short, I don't think we can argue that nonbaryonic cold dark
matter has proved to be considerably more stable than the
phlogistical chemistry that occupied much of Henry Cavendish's
scientific career [...]. (Peebles 1999
\cite{peebles99}).\footnote{Peebles has remarked, however, 
that he is somewhat more optimistic
regarding cold dark matter after the Boomerang and
Maxima results which were mentioned above (private communication).} 
\end{quote}

In addition to the nonbaryonic dark matter question, the usual
theoretical expectation from inflation is that $\Omega _{total}
=1$, and thus the total estimated matter content of the universe
is, even with new exotic forms of dark matter such as axions,
insufficient in relation to this (almost generic) inflation
prediction. So to make matter estimates compatible with inflation
expectations, particle cosmologists contemplate that a tiny part
of the vacuum energy which once drove the rapid expansion of the
universe may still be with us.\footnote{As indicated above, recent
observations of supernovae seem to require a small value of
$\Lambda$ ($\Omega_{\Lambda} \sim 2/3$), consistent with
$\Omega_{total}=1$, see e.g. \cite{turner99}.}

As with the baryon asymmetry, the question of the relationship
between cosmology and particle physics remains unanswered: while
cosmological observations can constrain the models suggested by
particle physics, none of the particle physics candidates for dark
matter have been clearly demonstrated in cosmological observations
or in particle accelerators.\footnote{There is some evidence that
one or more of the neutrinos may have a non-zero mass but, as
mentioned above, this is of little help since hot dark matter is
not believed to be the whole story.} The lack of evidence for a
particle physics origin of dark matter -- as well as the
conjecture of small amounts of vacuum energy of unknown source--
can therefore, according to taste, be regarded either as a problem
for the connection between cosmology and particle physics or as an
important motivation for further scrutinizing cosmological
observations and speculative particle physics models.

\ \\
{\em  Defining space and time}.
In modern cosmology the popular question ``what was before the Big
Bang?'' is mostly put aside as meaningless with the argument that
time and space were created with the Big Bang. But since the
principal scene for a unification between cosmology and particle
physics is the very {\em early} and very {\em small} universe,
reasonable notions of time and space are obviously required right
after the beginning for the unification to make sense. 

It is usually assumed that the backwards extrapolation of the
Friedmann equation becomes invalid at the Planck time, $t \sim
10^{-43}$ sec, where it is expected that quantum effects of
gravity will make classical general relativity break down. As
indicated earlier, theories of quantum gravity attempt to unify or
reconcile general relativity with quantum field theory but it is
still not clear how, or if, it is possible to achieve a theory
encompassing both Einstein's theory, where matter is treated
classically and spacetime is dynamical, and the quantum field
theory of matter which is formulated in a fixed spacetime
background.\footnote{Recent developments in quantum gravity
suggests that in order to perform this reconciliation, it might be
necessary to abandon the conventional ideas of quantum field
theory of depending on a metric structure \cite{rovelli99} -- but
is by no means clear how this would affect the successes of
standard QFT, which appear to be founded on the idea of fields
defined on a (fixed) spacetime background.} In spite of this, it
is sometimes assumed that quantization of the gravitational field
is, in fact, a necessity due to analogies with the famous 1933
Bohr-Rosenfeld analysis of the electromagnetic field quantization
(see e.g. \cite{kiefer99}). Rosenfeld has pointed out, however,
both that the analogy between the gravitational and the
electromagnetic field is problematic (due to the appearance of a
definite scale for space and time intervals in the theory of
gravity), and that the Bohr-Rosenfeld analysis showed only the
consistency of the electromagnetic field quantization, {\em not}
its necessity (\cite{rosenfeld63}, see also \cite{stachel99}
p.235). 

Although the treatments of space and time must be closely
connected in a relativistic framework, we shall here just comment
briefly of the possibilities of defining time in the ``early"
universe. Even if a theory of quantum gravity -- which presumably
governs the very early universe -- could be found, it is
notoriously difficult to see how the notion of time could somehow
emerge from such a theory. For instance, the canonical
Wheeler-DeWitt equation of quantum gravity does not depend on time
which obviously makes it hard to see if it can at all describe an
evolving universe (see e.g. \cite{kolb93} p.457). The problem of
how time might emerge from quantum gravity is discussed in detail
by, for example, Butterfield and Isham \cite{butterfield99a}.
These authors point out (p.57) that the emergence of time in
quantum gravity is not, in fact, a process in time, so that
contradictory questions such as `how did the universe look {\em
before} time emerged' can be avoided.

Nevertheless, even if quantum gravity can avoid downright
contradictions, there are at least two reasons to doubt whether
such a theory could be relevant for the description of our (early)
universe. First, although there might be no temporal relation
implied by the emergence of time from quantum gravity, it is still
highly uncertain how, and in what sense, time may be said to
emerge from quantum gravity -- not least because even
semi-classical approximations of the theory, where a time
parameter might be defined, are beset with technical and
conceptual problems (see e.g. \cite{butterfield99a} pp.49-52 and
references therein).\footnote{An aspect of these problems concerns
the difficulties involved in a straightforward application of
decoherence in the quantum cosmological context (as in e.g.
\cite{kiefer99} p.24-26). Specifically, it is difficult to see how
decoherence, which in standard quantum theory is construed as a
process in time, can be relevant to recovering the usual time concept
when the underlying quantum gravity theory is timeless
(\cite{butterfield99a} p.52).} Second, apart from this well-known
problem of how time can emerge in the theory, it is also hard to
make sense of the ``reverse" transition from time in the early
universe to a timeless quantum gravity. This problem, which could
be called {\em the reverse problem of time}, arises as follows:
One of the fundamental ingredients in the Big Bang scenario is the
existence of a cosmic time $t$ which is obviously necessary for 
discussing various epochs of the universe. The cosmic
time parameter is the proper time of a standard (classical) clock
at rest in the so-called comoving frame (see e.g. \cite{kolb94}),
and it is this time concept cosmologists have in mind when
discussing conditions in the {\em early} universe. Thus, the
assumption that quantum gravity (or quantum cosmology) is relevant
for the study of the very early universe rests on a solid
classical notion of time. But if it is conjectured that timeless
quantum gravity is {\underline {the}} fundamental theory, it
appears paradoxical that its central field of application (the
early universe) is only defined by a concept (classical cosmic
time) which is completely alien to the theory! In turn, if we
cannot even discuss the central application of quantum gravity
without assuming a classical time concept, it is unclear
what we should understand by an assumption like `the ultimate
nature of spacetime is non-classical' and, more generally, what we
should understand by the assumption that `classical physics can
ultimately be reduced to quantum gravity'. 
 
In connection with their discussion of quantum gravity  
and the `Copenhagen' interpretation of quantum mechanics, 
Butterfield and Isham note (\cite{butterfield99a} p.37): 
\begin{quote}
...we admit to giving this sort of interpretation little credence.
For we see no good arguments for the necessity of such
a background manifold [classical spacetime] as, for example, (in
Bohr's words) a `precondition of unambiguous communication'. 
\end{quote}
But, as we have just seen, even if it is assumed that quantum
gravity is relevant for cosmology, one can in Bohrian terms say
that a classical (cosmic) time indeed seems to be a necessary
condition for unambiguous discussions of the early
universe. 

These brief remarks can in no way do justice to the profound
conceptual problems with which a quantum theory of gravity is
faced, but -- like the cosmological constant problem issue -- they
illustrate that the search for a unified quantum gravity theory is
not necessarily a viable road to unity between cosmology and
particle physics.

\s{Summary and Conclusion}

So far, there is little or no observational evidence which
motivates the idea that particle physics is intimately related to
cosmology. Although cosmological data can constrain speculative
particle physics models which might explain features of the
present universe, these models have yet to receive direct
experimental or observational support -- either in physics
laboratories or in cosmological observations of clear footprints
from a particle physics dominated era in the early universe. Thus,
although the observed baryon asymmetry and the apparent need for
dark matter have led to interesting theoretical ideas beyond
standard particle physics, these ideas are still purely
speculative. And even the cosmological consequences of relatively
well established particle physics remain theoretical conjectures.
In particular, the observational pillars of the Big Bang provide
no direct support for the cosmology-particle physics connection --
simply because the scene for this connection is much earlier than
the epoch of nucleosynthesis, the release of microwave background
radiation, and the formation of galaxies which eventually have
permitted cosmologists to observe the Hubble expansion. Thus, it
has yet to be established that particle physics is {\em relevant}
to cosmology.

On the more theoretical side, we first reviewed the cosmological
constant problem which arises due to the assumed relation between
(physically real) quantum vacuum energy and general relativity.
While various types of solutions to this problem are conceivable,
we discussed some of the difficulties associated with construing
solutions in a unified framework of quantum gravity. In
particular, we noted with respect to quantum cosmology that 1) as
yet, there exists no consistent quantum theory of gravity and thus
no quantum cosmology, 2) dealing primarily with the very early
universe it would be difficult to test such a theory, and 3) it is
still far from clear that it makes sense to apply quantum
mechanics to everything in the universe or to the universe as a
whole. Given these general problems about imagining a solution of
the cosmological constant problem provided by a further unified
theory, the problem remains a source of doubt as to whether
general relativity and quantum field theory are at all compatible.
Consequently, the cosmological constant problem is a problem for
the {\em consistency} between cosmology and particle physics, at
least insofar as the latter is taken to involve physically real
vacuum energy.

As far as inflation is concerned, we have seen that this theory
does not necessarily provide optimism for an intimate connection
between cosmology and particle physics; for three reasons. 1) The
physical mechanism for inflation, as a rearrangement of the vacuum
structure in the early universe, is obscured by the cosmological
constant problem which tells us that the concept of vacuum is
still not well understood. 2) Inflation is epistemologically
doubtful: Since inflation only covers the first split second of
the universe it is difficult to test, and, conversely, the huge
variety of inflation models which can accommodate a wide range of
observational results makes it difficult to rule inflation out. 3)
While inflation was originally based on simple versions of Grand
Unified Theories and so was a ``good move" in the unification
scheme, we now know that inflation cannot be based on such
theories (which are also most likely wrong), and that some later
inflation models do not operate with quantum field-based inflation
ideas. It is thus no longer clear that a period of cosmological
inflation can be derived from fundamental physics.

As a last theoretical issue we have mentioned -- although by no
means discussed in detail -- that the differences in the treatment
of spacetime in quantum field theory and general relativity, as
well as the problematic notion of time in prototype versions of
quantum gravity, could mute optimism for finding a unified theory
in which cosmology and particle physics would be treated in a
common framework. This point makes it doubtful whether cosmology could
eventually be {\em derived} from particle physics in some
appropriate limit of quantum gravity.

With hesitations about both the relevance of particle physics to
cosmology, the consistency between GR and QFT, and the possibility
of deriving cosmology from a deeper quantum (gravity) theory --
let us return to the question of what kind of unity can be claimed
for particle physics and cosmology. It seems clear that {\em
ontological unity}, in which objects and phenomena in particle
physics and cosmology are interconnected, remains speculative.
Even the more so, if the interconnection between cosmology and
particle physics is conceived of in {\em reductionist} terms (i.e.
if cosmology is imagined to be reducible to particle physics).

It could perhaps be objected that these hesitations are premature,
for instance because the assumption of ontological unity between
cosmology and particle physics has already lead to a number of
interesting ideas which surely support the {\em methodological}
impact of such a conjecture. Indeed, the problems discussed in
this article might eventually provide fuel for new physics or new
observations which could support the unity between cosmology and
particle physics. Of course, nobody knows what the future will
bring; but it should be noted that a methodological strategy of
searching for unity -- like other guiding principles -- carries
with it the danger of neglecting alternatives to programmes
confined to this strategy. A worry along similar lines has been
expressed by Burbidge:

\begin{quote}
...the {\em beliefs} developed over the last twenty years in this
area [cosmology] have led to a total imbalance in the direction of
research. [...]  Far-out theoretical ideas are only
taken seriously, if they are related to the very early universe.
There, anything and everything goes. We are told that the unity
between particle physics and astrophysics is upon us. Suppose
there was no initial dense state? Suppose that matter and
radiation were never strongly coupled together? Suppose that the
laws of physics have evolved, as have everything else? The current
climate of opinion requires that these questions {\em not} be
asked. \cite{burbidge88}
\end{quote}
One does not have to subscribe to Burbidge's alternative `quasi
steady-state' cosmology \cite{burbidge99}, or to his apparent
sympathy for evolving laws of nature, to see that his
questions are relevant to further critical philosophical analysis
of the relationship between cosmology and particle physics.

On the other hand, if one takes seriously these challenges to
unity between cosmology and particle physics, one might very well
ask if I have anything to suggest instead. Without attempting a
full analysis there seems to be at least two options. 1) Perhaps
gravity is not a quantized force. As I have indicated, Bohr's
anti-reductionist notion of unity, in which classical concepts are
indispensable in quantum descriptions, is related to a more
pessimistic view on the prospects of quantum gravity: If classical
concepts (e.g. of time) are indeed necessary to discuss the early
universe then it seems impossible to understand the universe of
today solely as a consequence of a quantum gravity era, to the
extent that this would be an attempt to understand the universe in
terms of a `quantum' reality.\footnote{Insofar as matter and
radiation in the early Big Bang universe are assumed to be
described exclusively in quantum (particle physics) terms,
arguments of the in principle necessity of classical physics would
also be relevant for a critical discussion of cosmology without
quantum gravity.} If the gravitational force is classical, it
would mean that there is no quantum basis for understanding
general relativity and thus no common unified framework for
cosmology and particle physics. In this case, there can still be
theoretical bridges between micro- and macro-physics, and the
question of whether particle physics is relevant to cosmology is
therefore left open. 2) Perhaps we should take a more critical
approach to Big Bang; which, as Burbidge hints, is a central
source of dreams of unity between cosmology and particle physics
(given the supposed extreme energy conditions in the early
universe). Of course, one might argue that even if the present
understanding of the cosmological constant, inflation, dark matter
etc. is all wrong, there is no need to question the overall Big
Bang scenario. Nevertheless, if the standard picture is accepted
until, say, $10^{-11} s$ after the Big Bang, it is hard to resist
the temptation of extrapolating the Friedmann equation further
back in time in search of the initial conditions of our universe.
Thus, to really pursue this line of thinking, one would probably
have to doubt some of the central pillars in the Big Bang theory.

Whatever the fate of these alternatives, it is fair to say that
the connections between particle physics and cosmology, though
often cherished, do not yet warrant a claim for a successful case
of the unity of science. Again, this is not to say that the search
for unity should not be pursued. It is just that the result should
not be taken for granted.

\subsubsection*{Acknowledgements}
I would like to thank Svend E. Rugh for many discussions related
to the present article. It is also a pleasure to thank Jeremy
Butterfield, Mart\' \i n L\' opez Corredoira, Jan Faye, Helge
Kragh, Mariano Moles, and Jes\' us Moster\' \i n for comments on
earlier drafts of this paper. Financial support from EU contract
HPMF-CT-1999-00336 is gratefully acknowledged.

\end{document}